\documentstyle[aps]{revtex}

\def\beq{\begin{equation}}
\def\eeq{\end{equation}}
\def\bea{\begin{eqnarray}}
\def\eea{\end{eqnarray}}

\begin{document}

\draft

\title{Damping of a Yukawa Fermion at Finite Temperature}
\author{M.H. Thoma}
\address{Institut f\"ur Theoretische Physik, Universit\"at Giessen,\\
35392 Giessen, Germany}
\date{\today}
\maketitle

\begin {abstract}

The damping of a massless fermion coupled to a massless scalar particle
at finite temperature is considered using the Braaten-Pisarski resummation
technique. First the hard thermal loop diagrams of this theory are extracted
and effective Green's functions are constructed. Using these effective
Green's functions the damping rate of a soft Yukawa fermion is
calculated. This rate provides the most simple example for the damping of a
soft particle. To leading order it is proportional to $g^2T$, whereas the
one of a hard fermion is of higher order.

\end{abstract}



\section{Introduction}

A consistent calculation, i.e. leading to gauge independent results that
are complete in the order of the coupling constant, of damping rates in
quantum field theories at high temperature requires the use of resumed
propagators and vertices according to the effective perturbation theory
developed by Braaten and Pisarski \cite{r1}. Using this method the damping
rates of quarks and gluons in a quark-gluon plasma have been discussed in
great detail
\cite{r2,r6,r7,r8,r9,r10,r11,r12,r14,r15,r17,r18,r19,r20,r21,r22,r23,r24}.
For example, the damping rates of soft partons, i.e.
of partons with momenta of the order of $gT$ or smaller, were found to be
finite and proportional to $g^2T$ \cite{r21,r22,r23,r24}.
In particular, the long standing
puzzle of the gauge dependence of the gluon damping rate at rest was
solved in this way \cite{r21}. However, the computation of soft rates is
rather cumbersome since effective propagators as well as vertices have to be
taken into account.

In the case of the damping rates of hard partons with momenta
of the order of $T$ or
larger, on the other hand, it is sufficient to include only an effective
gluon propagator. These rates represent the most simple application of
the Braaten-Pisarski method in QCD. Therefore they have been considered in
a number of papers for studying the resummation technique, e.g. the
gauge independence of its results \cite{r11,r23,r25,r26}.
Furthermore these rates are
closely related to interesting properties of the quark-gluon plasma, such
as thermalization time, viscosity, and mean free path and energy loss
of partons \cite{r6,r19,r28,r29,r30,r31}.
However, owing to the absence of static magnetic
screening in the resumed gluon propagator the damping rate of hard partons
turns out to be logarithmic infrared divergent even using the resummation
technique \cite{r2,r6,r9,r17}.

Here we will investigate the case of a massless fermion coupled to a
massless scalar particle. The damping rate of this "Yukawa" fermion is of
interest because it provides a simple example for the application of the
Braaten-Pisarski method. Furthermore it shows some interesting features which
are not observed in gauge theories.

\section{Hard Thermal Loops and Effective Green Functions}

We start from the following Lagrangian
\beq
{\cal L} = i\> \bar \psi \> \gamma ^\mu \> \partial _\mu \psi + \frac {1}{2}\>
\partial _\mu \phi \> \partial ^\mu \phi + g\> \bar \psi \> \psi \> \phi\; ,
\label{e1}
\eeq
describing the coupling of a massless fermion $\psi $ to a massless scalar
$\phi$ with a coupling constant $g$.

According to the Braaten-Pisarski technique we first have to extract the
hard thermal loop diagrams of this theory from which we construct the
effective Green's functions by resummation. Starting with the hard thermal
loop self energy of the scalar particle we consider the diagram shown in
Fig.1. Standard Feynman rules give ($K^2=k_0^2-k^2$)
\beq
\Pi (P) = -i\> g^2\> \int \frac  {d^4K}{(2\pi )^4}\> tr\, [S(K-P)\> S(K)]\; .
\label{e2}
\eeq
Evaluating the trace over the gamma matrices and adopting the imaginary
time formalism at finite temperature, we find
\beq
\Pi (P) = -4\> g^2\> T\> \sum _{k_0} \int \frac {d^3k}{(2\pi )^3}\>
(k_0\, q_0-{\bf k}\cdot {\bf q})\> \widetilde \Delta (K)\> \widetilde
\Delta (Q)\; ,
\label{e3}
\eeq
where $Q=P-K$, $\widetilde \Delta (K)=1/K^2$ and the sum extends over discrete
fermionic energies, $k_0=(2n+1)i\pi T$. The sum over $k_0$ can easily be
performed by introducing the Saclay representation of the propagators
\cite{r32}
\beq
\widetilde \Delta (K) = -\int _0^{1/T} d\tau \> e^{k_0\tau }\> \frac {1}{2k}
\> \left \{[1-n_F(k)]\> e^{-k\tau } - n_F(k)\> e^{k\tau }\right \}\; ,
\label{e4}
\eeq
where $n_F(k)=1/[\exp (k/T)+1]$ is the Fermi distribution.

The calculation of the scalar self energy in the hard thermal loop
approximation, $K\gg P$, can be done analogously to the gluon polarization
tensor \cite{r1}. The only difference between (\ref{e3}) and the
corresponding expression for the longitudinal part of the gluon self
energy by quark polarization is, apart from an overall factor, a different
sign between the $k_0\, q_0$- and ${\bf k}\cdot {\bf q}$-term in
(\ref{e4}) coming from additional gamma matrices due to the quark-gluon
vertices in the trace of the gluon polarization tensor. This sign, however, is
essential since the minus sign in (\ref{e3}) leads to a cancellation of
the momentum dependent terms, from which the simple result

\beq
\Pi (P) = \frac {g^2T^2}{6}\; .
\label{e5}
\eeq
follows, whereas in the gluon case a complicated momentum dependent term
survives. Furthermore the gluon hard thermal loop polarization tensor
shows an imaginary part corresponding to Landau damping below the light cone.

The effective scalar propagator is given by resuming the hard thermal
loop  self energy (\ref{e5}) within a Dyson-Schwinger equation  resulting in
\beq
\Delta ^\star (K) = \frac {1}{K^2-m_S^2}\; ,
\label{e6}
\eeq
where $m_S^2=g^2T^2/6$ describes an effective thermal mass of the scalar
particle generated by the interaction with the fermions of the heat bath.

The hard thermal loop fermion self energy caused by the diagram of Fig.2
differs from the quark self energy \cite{r33}
only by a factor 8/3. Thus we
simply may replace the effective quark mass $m_q^2=g^2T^2/6$ by the effective
Yukawa fermion mass $m_Y=g^2T^2/16$. The effective fermion propagator in
the helicity eigenstate representation is given by \cite{r23,r34,r35}
\beq
S^\star (P) = \frac {1}{2D_+(P)}\> (\gamma _0 - \hat {\mbox{\boldmath $p$}}
\cdot \mbox{\boldmath $\gamma $}) + \frac {1}{2D_-(P)}\> (\gamma _0 +
\hat {\mbox{\boldmath $p$}} \cdot \mbox{\boldmath $\gamma $})\; ,
\label{e7}
\eeq
where
\beq
D_\pm (P) = -p_0\pm p+\frac {m_Y^2}{p}\> \left (\pm 1-\frac {\pm p_0-p}{2p}\>
\ln \frac {p_0+p}{p_0-p} \right )\; .
\label{e8}
\eeq
In contrast to the effective scalar propagator (\ref{e6}) the effective
fermion propagator (\ref{e7}) shows an imaginary part giving rise to
damping effects, as we will see below.

The Yukawa theory (\ref{e1}) contains no effective vertices on the hard
thermal loop level. Consider for instance the correction to the
three-point vertex shown in Fig.3. According to the rules of power
counting \cite{r1} this correction is suppressed compared to the bare
vertex by a factor of $g$ even if all external legs are soft. The reason for
this is the fact that the scalar propagator in Fig.3 is cancelled
by a factor $K^2$ coming from the fermion propagators in the hard thermal
loop limit. In QCD there is an additional gamma matrix in the vertex
correction, $K\! \! \! \! /\, \gamma _\mu \, K\! \! \! \! /
=2K_\mu \; K\! \! \! \!/
-K^2\, \gamma _\mu $. Owing to the first term the cancellation of the gluon
propagator is upset \cite{r1}. Similar considerations apply to higher
n-point functions. The absence of effective vertices is not surprising
as there are no Ward identities in the Yukawa theory, which relate
effective propagators to effective vertices.

\section{Damping of a Soft Yukawa Fermion}

Now we turn to the damping of a Yukawa fermion. First we consider the damping
of a hard fermion with momentum of the order of $T$ or larger. Comparing
with the quark damping rate the lowest order contribution proportional
to $g^2T$
\cite{r2,r6,r7,r8,r9,r10,r11,r12,r14,r15,r17,r18,r19,r20}
should come from the imaginary part of the fermion
self energy containing an effective scalar propagator (Fig.4). However, since
the scalar propagator has no imaginary part (virtual Landau damping)
the diagram of Fig.4 is real. The fermion self energy diagram containing
an effective fermion propagator, but a bare scalar propagator due to energy
momentum conservation at the vertex, contributes to higher order $g^4$, as
can be seen from the similar case of photon damping
\cite{r36}. Thus there is no damping of a hard fermion to order $g^2$
in the Yukawa theory.

The damping rate of a soft Yukawa fermion follows from the diagram of
Fig.5, where we have to use an effective scalar as well as fermion
propagator, since the external momentum is soft, $p_0$, $p\sim gT$.
The corresponding self energy diagrams for soft quark or gluon damping
are much more involved \cite{r21,r22,r23,r24}
due to the presence of effective
momentum dependent vertices and the momentum dependence of the effective
gluon propagator. The damping mechanism corresponding to the fermion
self energy of Fig.5 can be seen by cutting the diagram \cite{r37}. Since the
fermion propagator has an imaginary part there is a cut-pole contribution
describing "Compton scattering".

The damping rates of a soft fermion and a plasmino, corresponding
to a negative helicity eigenstate, at rest, $p=0$, are equal and
given by \cite{r23}
\beq
\gamma _\pm (0) = -\frac {1}{8}\> tr\, [\gamma _0\> Im \Sigma ^\star
(m_Y+i \epsilon,0)]\; ,
\label{e9}
\eeq
where $\Sigma ^\star $ is given by Fig.5.
Substituting the effective scalar and fermion propagators (\ref{e6}) and
(\ref{e7}) into $\Sigma ^\star$ and evaluating the trace over the
gamma matrices yields
\beq
tr\, [\gamma _0\> \Sigma ^\star (P)] =-2\> g^2\> T\> \sum _{k_0}\> \int
\frac {d^3k}{(2\pi )^3}\> \Delta ^\star (K) \> \left [\frac {1}{D_+(Q)} +
\frac {1}{D_-(Q)}\right ]\; ,
\label{e10}
\eeq
where $Q=P-K$. The sum over the Matsubara frequencies can be performed
easily by using the Saclay method again. In the case of the effective scalar
propagator the Saclay representation reads
\beq
\Delta ^\star (K) = -\int _0^{1/T} d\tau \> e^{k_0\tau }\>  \frac
{1}{2\omega _k}\> \left \{ [1+n_B(\omega _k)]\> e^{-\omega _k\tau } +
n_B(\omega _k)\> e^{\omega _k\tau }\right \} \; ,
\label{e11}
\eeq
where $n_B(\omega _k)=1/[\exp (\omega _k/T)-1]$ denotes the Bose distribution
and $\omega _k^2=k^2+m_S^2$. In the case of the effective fermion propagator
it is convenient to introduce the spectral representation \cite{r38}
\beq
\frac {1}{D_\pm (Q)} = -\int _0^{1/T} d\tau '\> e^{q_0\tau '}\>
\int _{-\infty }^\infty d\omega \> \rho _\pm (\omega ,q)\>
[1-n_F(\omega )]\> e^{-\omega \tau '}\; ,
\label{e12}
\eeq
where the spectral functions $\rho _\pm $ are given in Ref.\cite{r34,r35}.
Since the damping comes from the cut term of the fermion propagator we only
need the discontinuous part of the spectral function defined by
\beq
\rho _\pm ^{disc} (x,y) = \frac {1}{\pi }\> Im\, \frac {1}{D_\pm
(x,y)}\> \theta (y^2-x^2)\;.
\label{e13}
\eeq

Carrying out the sum over $k_0$ and the integrations over $\tau $ and
$\tau '$ (\ref{e10}) reduces to
\bea
tr\, [\gamma _0\> \Sigma ^\star (P)] & = & g^2\> \int \frac
{d^3k}{(2\pi )^3}\> \frac {1}{\omega _k}\> \int _{-\infty }^\infty d\omega \>
[\rho _+^{disc}(\omega ,q) + \rho _-^{disc}(\omega ,q)]\nonumber \\
& \! & \left \{ [1-n_F(\omega )+n_B(\omega _k)]\> \frac {1}{p_0-\omega _k-
\omega } + [n_F(\omega )+n_B(\omega _k)]\> \frac {1}{p_0+\omega _k-\omega }
\right \} \; .
\label{e14}
\eea
Putting the self energy $\Sigma ^\star$ on the mass shell,
$p_0=m_Y+i\epsilon$, $p=0$, extracting the imaginary part via
\beq
Im\, \frac {1}{m_Y\pm \omega _k-\omega +i\epsilon} = -\pi \> \delta
(m_Y\pm \omega _k-\omega )\; ,
\label{e15}
\eeq
and expanding the distribution functions for soft energies, $n_F(\omega )
\simeq 1/2$ and $n_B(\omega _k)\simeq T/\omega _k$, we obtain
after integrating
over the angles and $\omega $ the leading term of the damping rate
\beq
\gamma _\pm (0) = \frac {g^2T}{16\pi }\> \int _0^\infty dk\> \frac {k^2}
{\omega_k^2}\>  [\rho _+^{disc}(\omega _+,k)+\rho _+^{disc}(\omega _-,k)
+\rho_-^{disc}(\omega _+,k)+\rho _-^{disc}(\omega _-,k)]\; ,
\label{e16}
\eeq
where $\omega _\pm=m_Y\pm \omega _k$.

Inserting (\ref{e13}) into this expression, the step function of the
discontinuous part of the spectral function cannot be satisfied for
$\omega _+$ which is always larger than $k$. In the case of $\omega _-$
the step function restricts $k$ to
\beq
k\geq \sqrt {\left (\frac {m_S^2+m_Y^2}{2m_Y}\right )^2-m_S^2}=\frac {5}{24}
\> g\> T\; .
\label{e17}
\eeq

Carrying out the remaining integration over $k$ numerically, we end up with
the final result
\beq
\gamma _\pm (0) = 0.24\> \frac {g^2T}{16\pi }\; .
\label{e18}
\eeq

\section{Conclusions}

We have considered the Yukawa theory at finite temperature. First we have
extracted the hard thermal loops and constructed effective Green's functions
from them. In contrast to gauge theories the boson propagator shows
no momentum dependence and is real for all values of the momentum and
the energy. Also there are no effective vertices in this theory.

Next we have calculated the damping rate of a Yukawa fermion. Owing to the
absence of an imaginary part of the effective scalar propagator the
damping rate of a hard fermion is at least of the order $g^4$. The
damping rate of a soft fermion, on the other hand, is of the order $g^2$
due to the simultaneous use of an effective scalar and fermion propagator.
As a new feature, not observed in the case of quarks or gluons, the
damping rate of a Yukawa fermion at rest contains a kinematic
restriction depending on the effective thermal masses of the scalar
field and the Yukawa fermion.

The computation of the Yukawa fermion damping rate at rest provides
the most simple example for a soft rate using the Braaten-Pisarski method
since no effective vertices are involved and the effective scalar
propagator is momentum independent.

\acknowledgements

We would like to thank R.D. Pisarski for drawing our attention
to this problem and helpful discussions.

\begin{figure}
\caption{Hard thermal loop contribution to the scalar self energy.}
\end{figure}

\begin{figure}
\caption{Hard thermal loop contribution to the fermion self energy.}
\end{figure}

\begin{figure}
\caption{Correction to the three point vertex.}
\end{figure}

\begin{figure}
\caption{Fermion self energy containing an effective scalar
propagator denoted by a blob.}
\end{figure}

\begin{figure}
\caption{Fermion self energy containing an effective scalar and fermion
propagator.}
\end{figure}

\end{document}